\documentclass[runningheads]{svmult}

\usepackage{makeidx}   
\usepackage{graphicx}  
\usepackage{subeqnar}  
\usepackage{multicol}  
\usepackage{physprbb}  
\makeindex             

\def\Fig#1{Fig.~\ref{fig:#1}}
\def\kms{\, \rm{km}\,  \rm{s}^{-1}}
\def\rvir{R_{\rm vir}} 
\def\vvir{V_{\rm vir}}
\def\lp{\lambda'}
\def\lpdm{\lambda'_{\rm dm}}
\def\lpb{\lambda'_{\rm b}}
\def\sglp{\sigma_{\lambda'}}

\def\rb{R_{\rm b}}
\def\rdm{R_{\rm dm}}

\def\jmax{j_{\rm max}}
\def\jtot{j_{\rm tot}}
\def\Jb{J_{\rm b}}
\def\Jdm{J_{\rm dm}}
\def\fbar{f_{\rm b}} 
\def\fd{f_{\rm d}} 
\def\vfb{V_{\rm fb}}
\def\vsat{V_{\rm sat}}

\begin{document}

\vskip -2cm
\title*{Resolving the Spin Crisis: Mergers and Feedback} 
\toctitle{Resolving the Spin Crisis: Mergers and Feedback}
\titlerunning{Resolving the Spin Crisis: Mergers and Feedback}
\author{Avishai Dekel\inst{1}
\and Ariyeh H. Maller\inst{1}
}
\authorrunning{Dekel \& Maller}
\institute{Racah Institute of Physics, The Hebrew University,
Jerusalem 91904, Israel
     }
     
\maketitle             


\begin{abstract} 
We model in simple terms the angular momentum ($J$) problem of galaxy formation
in CDM, and identify the key elements of a scenario that can solve it.  The 
buildup of $J$ is modeled via dynamical friction and tidal stripping in 
mergers.  This reveals how over-cooling in incoming halos leads to transfer of 
$J$ from baryons to dark matter (DM), in conflict with observations.  By 
incorporating a simple recipe of supernova feedback, we match the observed $J$ 
distribution in disks.  Gas removal from small incoming halos, which make the 
low-$J$ component of the product, eliminates the low-$J$ baryons.  Partial 
heating and puffing-up of the gas in larger incoming halos, combined with tidal
stripping, reduces the $J$ loss of baryons.  This implies a higher baryonic 
spin for lower mass halos.  The observed low baryonic fraction in dwarf 
galaxies is used to calibrate the characteristic velocity associated with 
supernova feedback, yielding $\vfb \sim 100\kms$, within the range of
theoretical expectations.  The model then reproduces the observed distribution 
of spin parameter among dwarf and bright galaxies, as well as the $J$ 
distribution inside these galaxies.  This suggests that the model captures the 
main features of a full scenario for resolving the spin crisis.
\end{abstract}

\section{Introduction}
\label{sec:intro}

The `standard' model of cosmology, CDM, which assumes hierarchical buildup of 
structure, is facing difficulties in explaining 
observed properties of galaxies, such as the number density of dwarfs
and the inner density profile of halos. 
Standing out is the angular-momentum problem,
that is the apparent failure of the theory, via simulations,
to reproduce the large sizes of disk galaxies and their structure.
We make progress by first reproducing the problem via a simple model in 
which the important physical elements are spelled out,
and then incorporating in this model the key process which may cure
the problem --- feedback.

The sizes of disks are commonly linked to the spins of their parent halos
as measured in N-body simulations [10]. 
The assumptions that the baryons and DM share the same
distribution of specific angular momentum $j$ and that the baryons conserve
their $j$ while contracting to a disk lead to disk sizes comparable to those
observed.  However, simulations that incorporate gas find that most of the
baryonic $j$ is transfered to the DM, resulting in disk sizes smaller by an
order of magnitude 
[e.g.~14,15],
and thus leading to a {\it spin catastrophe}.

In addition, there is a {\it mismatch of $j$ profiles}.
The $j$ distribution (or profile) within simulated halos has been found to
scatter about a universal shape, with an excess of low-$j$ (and high-$j$) 
material compared to the exponential disks observed [1, BD].
This mismatch has been demonstrated in an observed sample of 14 dwarf
galaxies [18, BBS],
which serves as the target for our modeling effort.
BBS used for each halo the measured rotation curve and an assumed NFW profile
to determine the halo virial quantities, with an average
$\langle \vvir \rangle \simeq 60\kms$. They then determined the baryonic
spin parameter, averaging to $\langle \lpb \rangle \sim 0.07$,
significantly larger than the $\langle \lpdm \rangle \sim 0.035$
of simulated halos, and then demonstrated the $j$-profile mismatch
case by case.  BBS also estimated the ratio of disk to DM mass to be
$\langle \fd \rangle \sim 0.04$, about a factor of 3 smaller than the
universal fraction, indicating significant gas loss.

The spin catastrophe is commonly being associated with ``over-cooling", that 
the gas rapidly cools and becomes tightly bound in small halos.
When such a halo spirals into a bigger halo, the baryonic component
survives intact all the way to the center and thus transfers
all its orbital $j$ to the DM.
It has therefore been speculated that energy feedback from
supernova may remedy the problem by balancing the early cooling
[e.g.~9].
A key idea is that the spin segregation between
baryons and DM can go either way.  While gas cooling tends
to lower the baryonic spin, heating due to feedback would reduce this effect, 
and gas removal from small halos would even lead to higher baryonic spin.
However, a realistic implementation of feedback has proved challenging
[e.g.~17].
The feedback process has not yet been studied or implemented
in satisfactory detail. We do not know yet whether they can indeed solve
the CDM problems, and how. 
This motivates our attempt to first understand how the feedback scenario
may work using a very simple semi-analytic model. 
Knowing that in a hierarchical scenario the halo fromation can be largely
interpreted as a sequence of mergers, our model is 
based on a simple algorithm for the buildup of halo spin by adding up the 
orbital angular momenta of merging satellites 
[13, MDS; 19].
It matches well the spin distribution among halos in N-body simulations
as well as the $j$ profile within halos.
This makes it a useful tool for understanding the over-cooling origin of 
the spin problem and for the attempt to cure it via feedback effects.
Our work is described in more detail in 
[12, MD].

\section{Buildup of Halo Spin by Mergers}
\label{sec:halos}

We characterize the angular momentum $J$ of a galaxy by the modified spin
parameter {bull:01} $\lp=(J/M)/(\sqrt{2}\vvir\rvir)$.  
This quantity, which equals the standard $\lambda$
for an isothermal sphere and for an NFW profile with concentration $c\sim 5$,
is straightforward to compute separately for the baryons
and for the DM, $\lpb$ and $\lpdm$.
The distribution of $\ln \lp$ in the simulations is normal,
with an average corresponding to $\lp_0 \simeq 0.035$ 
(compared to $\lambda_0 \simeq 0.042$)
and a standard deviation $\sglp \simeq 0.5$.
The ``orbital-merger" model of MDS reproduces this spin distribution.
To materialize this model we generate many random realizations
of merger histories based on the Extended Press Schechter formalism
[11]
with slight adjustments,  
and for each merger tree we create random realizations of the orbital ${\bf J}$
added in each merger. The encounter parameters are taken 
to mimic typical mergers, with the directions of the orbits
drawn at random (or fine-tuned for a slight correlation between 
successive mergers as seen in simulations 
[5,16].
The resultant distribution of halo spins matches the
log-normal distribution obtained in the simulations. 

The cumulative mass distribution of $j$ within simulated halos is fit 
by the universal function $M(<j) = M_{\rm v} {\mu\, j}/(j_0 + j)$,
with $\mu > 1$ and $j \leq j_{\rm max} = j_0/(\mu - 1)$ (BD).
This is a simple power law, $M(<j) \propto j$, for at least half
the mass, with a possible bend characterized by $\mu$.
The other parameter, $j_0$ or $\jmax$, can be replaced by $\lp$.
The distribution of $\mu$ is Gaussian in $\ln{(\mu-1)}$, with a mean $-0.6$.
The model also recovers these simulated $j$ profiles.
We create an $M(<j)$ profile for each of the EPS model realizations 
by dividing the mass growth of the halo into bins and assigning to each
the corresponding orbital $J$.
A sample of profiles produced by this procedure and the distribution of 
$\mu$ values are shown in Figs.~1 and 2 of MD,
demonstrating the match with the simulation results of BD.
The model also reproduces the insensitivity to halo mass and redshift.
 
The successes of the simple model in recovering both
the distribution of spins and the $j$ profiles 
makes it a useful tool for studying the $j$ buildup.
A new feature revealed by the model, which provides an interesting clue, is 
that the final halo spin is predominantly determined by the last major merger, 
while the many smaller satellites come in at different directions and therefore 
tend to sum up to a low $j$.  If small satellites would lose 
gas before they merge into the halo, then much of the galactic gas
would originate in big satellites, the final gas  would lack the low-$j$ 
component, and the baryonic spin would end up higher than the DM.

\section{Reproducing the Baryonic Spin Loss}
\label{sec:oc}

We can understand the $j$ loss of baryons via a simple 
model including gas cooling, dynamical friction and tidal stripping
for how the orbital $j$ is converted into halo spin.  First, the dynamical 
friction exerted by the halo on the satellite brings the satellite 
towards the halo center and thus transfers $j$ from the orbit to the halo. 
Second, once satellite particles are tidally stripped they retain their $j$ 
at the stripping point and add it directly to the halo. 

The mass loss at halo radius $r$ can be estimated by evaluating 
the tidal radius $\ell_{\rm t}$ of the satellite at $r$ via the resonance 
condition, ${m(\ell_{\rm t}) / \ell_{\rm t}^3}={M(r) / r^3}$,
where $m(\ell)$ and $M(r)$ are the mass profiles of the satellite and halo.
If these two are self-similar, then this implies
$\ell_{\rm t}/\ell_{\rm vir} =r/\rvir$, 
and the bound mass of the satellite is 
$m[\ell_{\rm t}(r)] \propto M(r)$.
A more accurate recipe for tidal stripping, tested with merger simulations, 
reveals that this is a good approximation in general 
[6],
so we adopt it in our model.

\begin{figure}[] 
\begin{center}
\vskip 4.3 truecm
\includegraphics{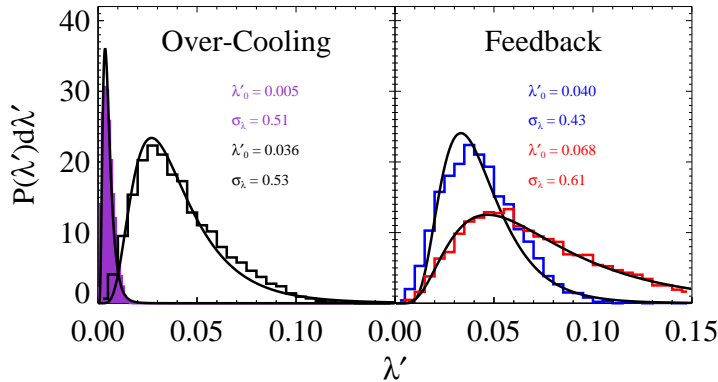}
\end{center}
\caption[]
{The effects of over-cooling and feedback on the spin distribution
of baryons compared to the DM, for dwarf and bright galaxies.
Log-normal fits are shown, with the mean and scatter quoted.
{\bf Left:} $\lpb$ is shifted down by an order of magnitude
compared to $\lpdm$, reproducing the spin catastrophe.
{\bf Right:} $\lpb$ in bright galaxies is boosted up by heating and partial 
blowout in incoming halos and roughly matches $\lpdm$, while in dwarf 
galaxies $\lpb$ is boosted up further by the blowout in small satellites.}
\label{fig:oc}
\end{figure}

When exploring the effect of cooling, one assumes that initially the baryons 
follow the DM.  As the gas cools, it contracts to a more compact configuration 
of radius $\rb<\rdm$.
This spatial segregation in the satellite implies that the $j$-rich mass
stripped at the early stages of the merger in the outer halo is dominated
by DM, while the more compact baryons penetrate into the inner halo
and lose more of their $j$ via dynamical friction.
The result is a net spin transfer from the baryons to the DM.
Using the stripping recipe $m(r)\propto M(r)$, 
we obtain for the final baryonic spin $\Jb/\Jdm =(\rb/\rdm)$. 
In the case of maximum cooling, the baryons dominate the halo center, 
$\rb = \fbar\rdm$, where $\fbar \simeq 0.13$ is the universal baryon fraction.
\Fig{oc} shows the resultant baryonic spin distribution according to this 
model; there is a shift down to $\lp_0=0.005$, reproducing the spin crisis.
The role of feedback would be to delay the cooling, increase $\rb$,
and thus reduce the baryonic spin loss.

\section{Feedback} 
\label{sec:fb}

Our approach here is to avoid the details of star formation 
and feedback and rather appeal to a very simple prescription for the 
effect of feedback as a function of the satellite's virial velocity, $\vsat$. 
Following the analysis of Dekel \& Silk 
[8],
we assume that the feedback
by supernova-driven winds pumps energy into the gas and heats it uniformly
to a temperature corresponding to a characteristic velocity $\vfb$,
on the order of $100 \kms$.  We therefore assume that 
the spatial extent of the baryons is determined by the 
ratio $\vfb/\vsat$.
The limit $\vsat \gg \vfb$, of massive, deep potential wells,
corresponds to maximum cooling, $\rb \ll \rdm$.  In smaller halos, Still 
with $\vsat \simeq \vfb$, we expect the heating to balance the cooling and
yield $\rb \simeq \rdm$. Our model is therefore an interpolation between these 
limits, $\rb=(\vfb/\vsat)^{\gamma_1}\rdm$, with $\gamma_1$ an arbitrary 
exponent, which we set for now to be unity.

If $\vfb$ is larger than $\vsat$, the feedback can cause gas blowout.
We assume that partial blowout starts occurring once $\vfb$ 
exceeds the escape velocity of the satellite,  
$\sim \sqrt{2}\vsat$, while total blowout is expected for $\vfb \gg \vsat$.  
We therefore parameterize the amount of gas that remains in the halo by
another interpolation, $\fd=(\vfb/\vsat/\sqrt{2})^{\gamma_2}$, 
with $\gamma_2$ an arbitrary exponent tentatively set to unity.
We report here the results for the simplest choice $\gamma_1=\gamma_2=1$, and  
explore the robustness of the results to different 
choices of $\gamma_1$ and $\gamma_2$ in MD.  

In \Fig{oc} we demonstrate the effects 
of this feedback scheme, with $\vfb=95 \kms$, on the 
distribution of $\lpb$. We do it for two kinds of final halos,
with $\vvir =60$ and $220\kms$, representing {\it dwarf\,} and
{\it bright\,} galaxies.
The baryons in the bright galaxies end up with spins comparable 
to their DM halos, with $\lp_0=0.042$, while in dwarfs they 
have significantly higher spins, with $\lp_0=0.067$. 
We learn that $\lpb$ in dwarfs, which are build up by
small satellites, is dominated by the blowout from these
satellites, and it ends up with $\lpb > \lpdm$. For bigger galaxies,
which are largely made of bigger satellites, the dominant effect 
is the heating, with some contribution from blowout, together leading
to a $\lpb$ distribution similar to $\lpdm$, 
in general agreement with observations.

In MD we explore a range of values for the exponents $\gamma_1$ and $\gamma_2$.
For each choice we determine $\vfb$ such that for the dwarfs 
$\langle \fd \rangle = 0.04$ as in BBS.
We find that our results for dwarfs remain practically unchanged,
while the results for bright galaxies have a weak dependence on the value of 
$\gamma_1$ in the range $(0.5,3)$.

\section{Model versus Observations}
\label{sec:obs}

\begin{figure}[] 
\begin{center}
\vskip 2.5 truecm
\includegraphics{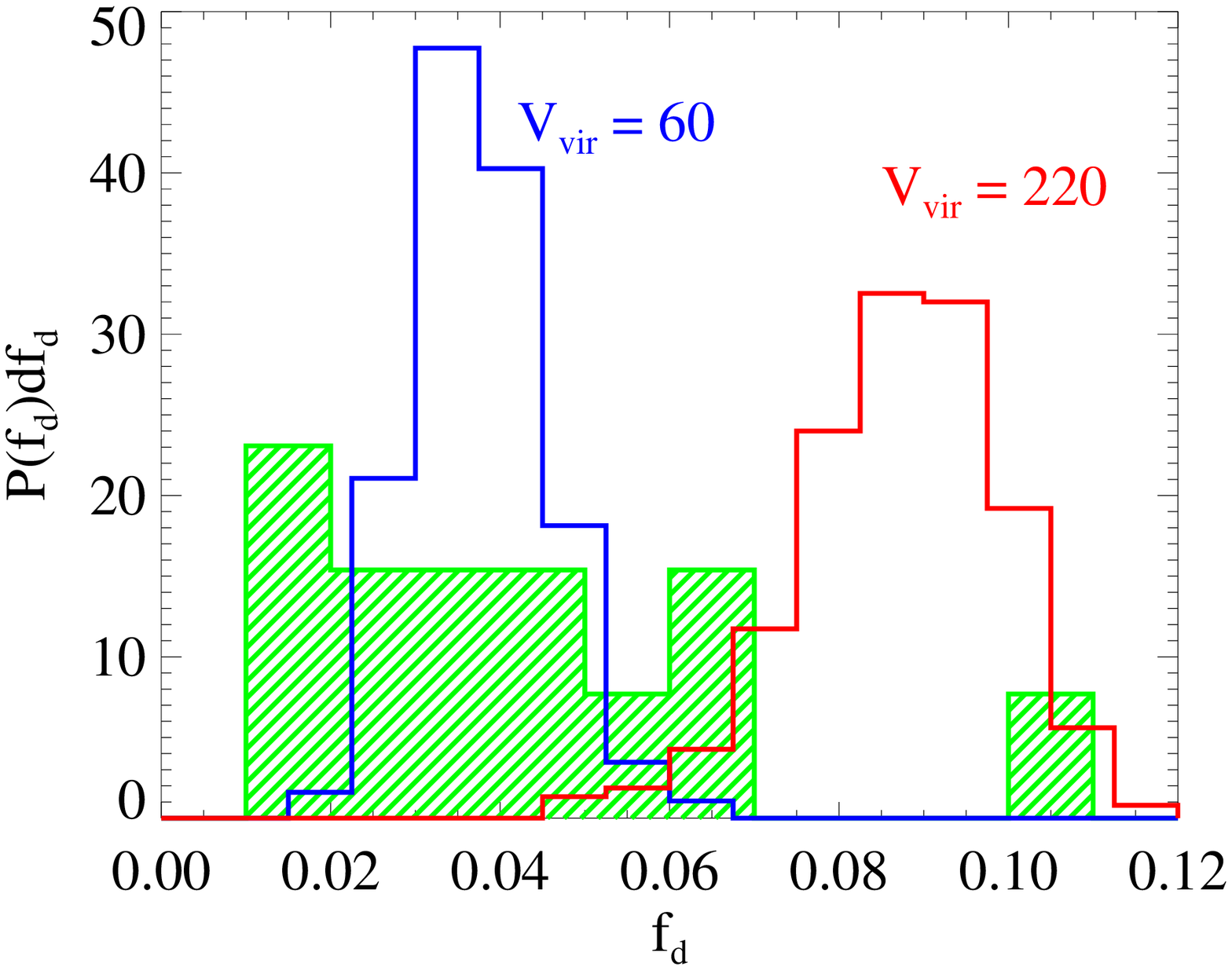}
\includegraphics{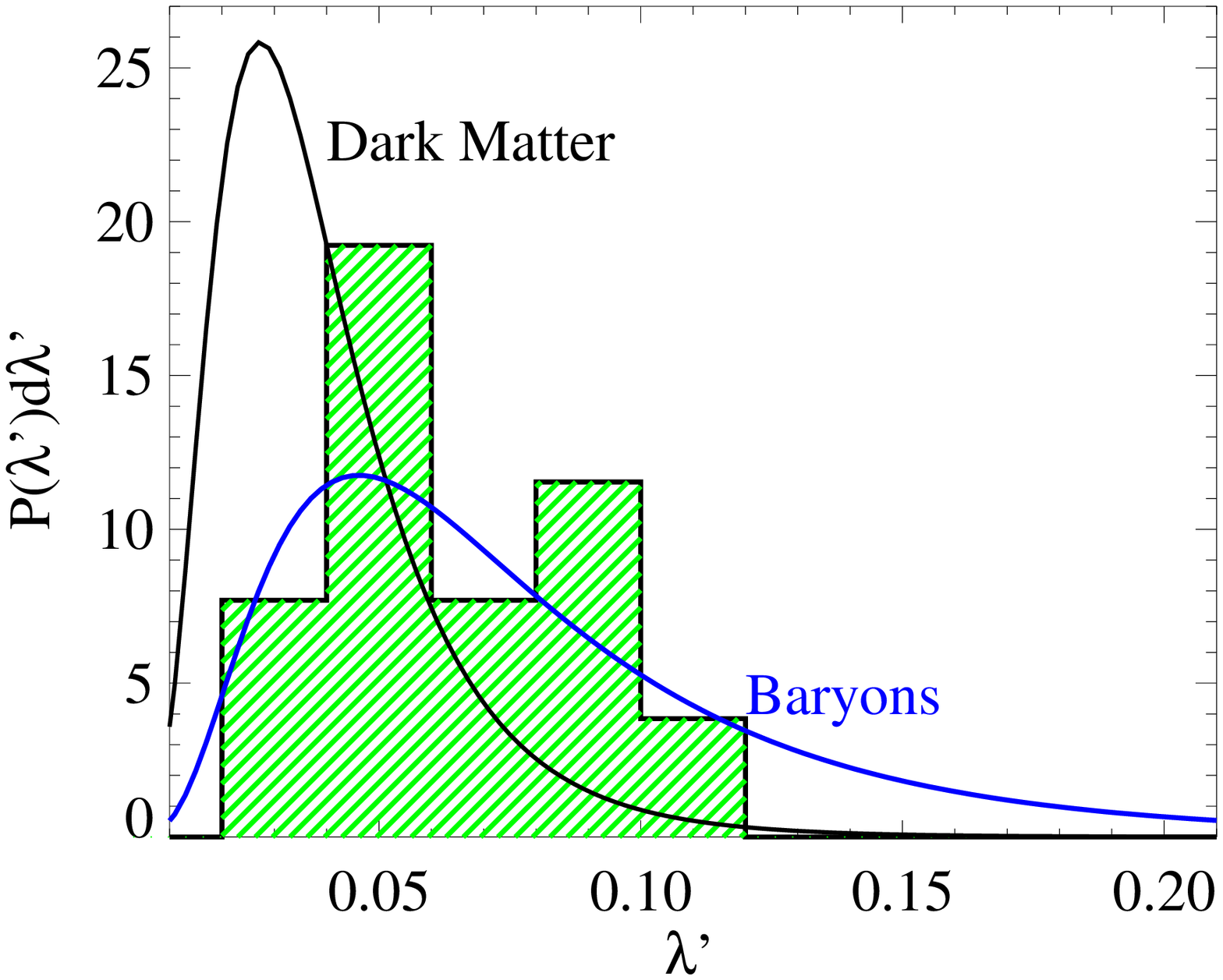}
\includegraphics{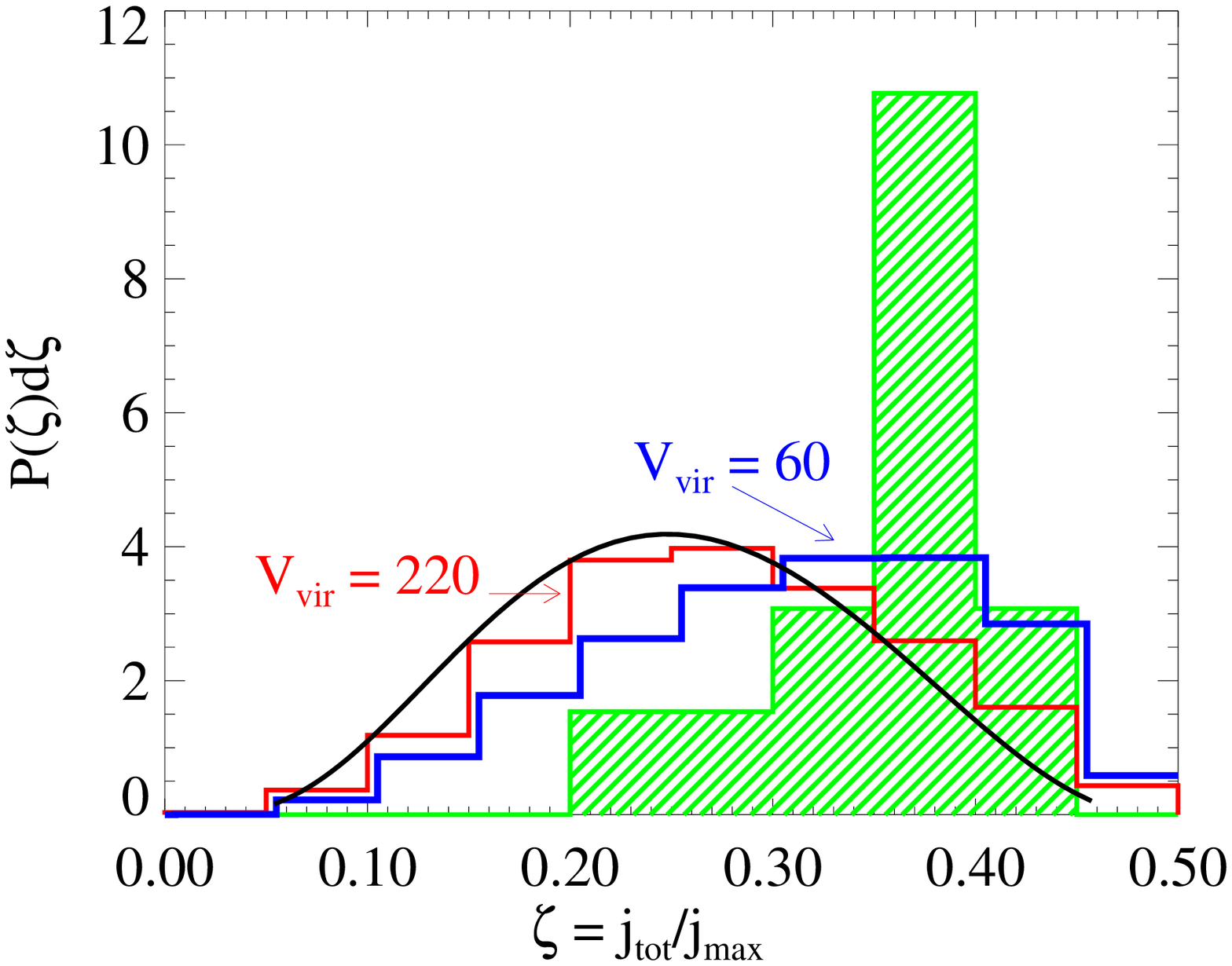}
\end{center}
\caption[]{
Model realizations (with $\vfb=95\kms$) 
versus BBS observations of dwarf galaxies (shaded).
Distribution of baryon fraction $\fd$, spin parameter $\lp$,
and $j$-profile shape parameter $\zeta$.
{\bf Left:} 
The model prediction for $\vvir=60\kms$ dwarfs, with significant blowout,
is in agreement with the BBS data, while the
bright galaxies retain most of their baryons.
{\bf Middle:} 
The predicted $\lpb$ distribution is in agreement with the dwarf data.
Shown for comparison is the simulation result for dark halos, which is
similar to that of bright galaxies.
{\bf Right:}
The predicted distribution of $\zeta$ for the baryons in dwarfs (heavy
histogram) is shifted upwards compared to the DM (smooth curve) and 
the bright galaxies (light histogram), like the data, but its width 
is overestimated.
}
\label{fig:fd}
\end{figure}

The distribution of $\fd$ for the dwarfs observed by BBS is displayed in
\Fig{fd}, showing values significantly lower than the
universal value of $\fbar \simeq 0.13$ and thus
consistent with baryonic blowout.
Shown for comparison are the model predictions for
dwarf and bright galaxies. We enforced a match of the means for the dwarfs at
$\langle \fd \rangle = 0.04$ by choosing $\vfb = 95 \kms$,
but the scatters are also in agreement.
For bright galaxies, $\fd$ is typically lower than the universal value
by less than $50\%$, reflecting the limited fraction of small progenitors
who lost their gas.

Next, we compare predicted and observed spin distributions for 
dwarfs.  We convert each value of $\lambda$ as quoted by BBS to $\lp$,
and show their distribution in \Fig{fd}. the observed spins are significantly
higher than the $\lp$ values of halos in cosmological simulations,
with an average of $\lp_0 \simeq 0.07$ compared to $0.035$. 
Then shown is our model prediction with $\vfb=95\kms$ for the 
baryonic spin distribution in dwarfs.
The effect of blowout brings the baryonic spins to a good 
agreement with the observed dwarfs.

\begin{figure}[] 
\begin{center}
\vskip 4.2 truecm
\includegraphics{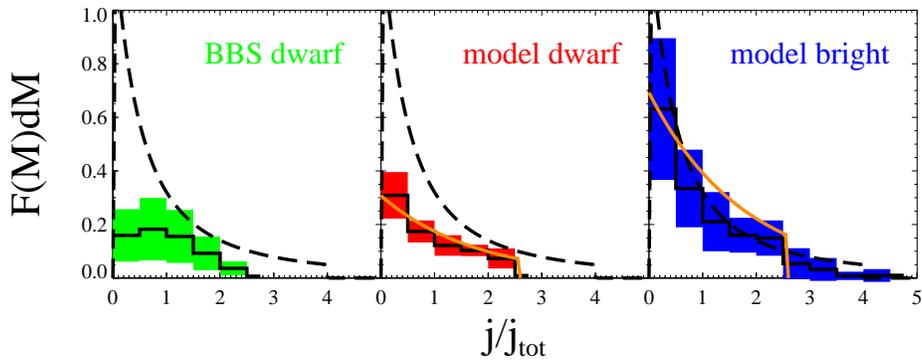}
\end{center}
\caption[]{
Average $j$ profiles (histogram) and the $1\sigma$ scatter (shaded)
for the observed dwarfs in comparison with the model dwarf and bright
galaxies ($\vfb=95\kms$). The integral under the histogram is $\fd$.
Shown in comparison is the typical profile for DM halos by BD. 
}
\label{fig:amd60}
\end{figure}

\Fig{amd60} shows the average $j$ profiles and the scatter about them
for the observed BBS dwarfs and for the corresponding model realizations
compared to the typical $j$ profile in halos by BD.
We construct the baryonic $j$ profile in each of our model realizations
following the same method used to produce DM $j$ profiles 
but now including feedback effects.
The BBS dwarfs show low baryonic fractions (indicated by the integral under
the histogram) and significant deficits of $j$
at the two ends of the distribution compared to the halos. 
The profile for model dwarf galaxies is similar to the observations
except for the very lowest $j$ bin which is a $2\sigma$ overestimate,
representing a spike in some of the model realizations. 
This spike may correspond to a low-$j$ baryonic component that
BBS fail to observe (faint halo stars?), or the spiky objects 
do not become disk dwarf galaxies, or our model needs to be improved.
The high-$j$ tail tends to be reduced in the baryons 
because it is often the result of a small satellite that comes in with 
its orbital $J$ aligned with the halo spin, and now has lost its gas.

\Fig{amd60} also shows the average model prediction for bright galaxies,
$\vvir=220\kms$.  They retain most of their baryons, so their profiles are 
less affected by blowout.  The average model profile is in better agreement
with an exponential disk,
towards solving the $j$-profile discrepancy pointed our by BD. 

A quantity used by BBS to characterize the shape of the $j$ profile,
as an alternative to the BD parameter $\mu$, is $\zeta \equiv \jtot/\jmax$.
In \Fig{fd} we also plot the distribution of this quantity in the BBS dataset 
in comparison with our model predictions for the baryons in dwarf
and bright galaxies. The predicted $\zeta$ distribution for dwarfs
is shifted upwards compared to the halos and the bright galaxies,
in qualitative agreement with the BBS data, but the width is overestimated.

\section{Conclusion} 
\label{sec:conc}

We devised a simple model to address the $j$ problems of galaxy formation 
within CDM.  By adding up the orbital ${\bf J}$ in random realizations 
of merger histories, the model successfully reproduces the simulated 
distribution of spins among halos (MDS) and the distribution of $j$ within 
halos (MD).  A simple analysis of how the merger orbital $j$ turns into
a spin profile provides a clue for how feedback effects in the
satellite can resolve the spin problems.
The idea is that the effective size of the gas component within the 
incoming halo determines its tidal stripping position in the big halo
and thus its final remaining baryonic spin after the merger.
The finding that the low-$j$ material originates in many minor mergers, that
tend to cancel each other's ${\bf J}$, provides the clue for
a possible solution to the $j$-profile mismatch problem.
The blowout of gas from small incoming halos, which is more pronounced in
satellites of dwarfs, would eliminate the low-$j$ baryons in the merger 
product and increase the spin parameter, as observed. 

The feedback effects, including heating and blowout, are modeled as a 
function of halo virial velocity, with one free parameter ---
the characteristic velocity $\vfb$ 
corresponding to the feedback energy from supernovae. 
To match the low baryonic fraction observed in dwarfs it has
to be $\vfb \sim 100 \kms$, consistent with the theoretical predictions 
[8].
This leads to an agreement between the model 
predictions and the observed disks, for the distribution of baryonic spin 
among galaxies and the baryonic $j$ distribution within galaxies, 
both dwarfs and bright galaxies.

We attempt to resolve the problems within the successful cosmological 
framework of CDM, by appealing to inevitable feedback effects.
Another approach is to appeal to the Warm Dark Matter (WDM) scenario,
despite the fact that it requires fine-tuning of the particle mass to 
$\simeq 1~keV$.  The main feature of WDM is the suppression of small halos
and the corresponding mergers.  While an N-body simulation of WDM 
[2]
indicates the same $j$ properties of halos
(the same properties can also be obtained as a general result
of tidal-torque theory, see MDS), one expects the cooling to be less
efficient in the absence of small halos, and thus the baryonic spin to be
higher. However, the $j$ profile is still expected to be a problem,
and the weaker feedback effects in the absence of small halos
may not be enough for resolving it.
These issues are yet to be studied in hydro simulations of WDM.

Feedback effects may also provide the cure to the missing dwarf problem in CDM,
where the predicted number of dwarf halos is much larger than the observed 
number of dwarf galaxies 
[3].
While the number of dwarfs is automatically suppressed in WDM,
it seems that the inclusion of the minimum inevitable feedback effects 
would reduce the predicted number of dwarfs to significantly below the 
observed number and thus be an overkill (J. Bullock, private comm.).
Finally, we find 
[7]
that the key 
elements of our toy model --- the tidal effects in mergers and the feedback 
in small halos --- are also very relevant in understanding and resolving the 
third problem of CDM, where the halos in simulations typically show steep 
cusps in their inner profiles 
[14],
while observations indicate flat cores at least in some galaxies
[4].
An analysis of tidal effects explains the inevitable formation of an asymptotic
cusp as long as satellites continue penetrating into the halo center. 
Feedback effects may puff up small satellites, make them disrupt in the outer
halo and thus allow a stable core.

The success of our toy model in matching several independent observations 
indicates that it indeed captures the relevant elements of the complex 
processes involved, and in particular that feedback effects may indeed 
provide the cure to all three problems of galaxy formation in CDM.
The next natural step should be to incorporate
a more sophisticated feedback recipe into the model using  
semi-analytic models and then full-scale cosmological simulations.


This research has been supported by the Israel Science Foundation
grant 546/98, by the US-Israel Binational Science Foundation
grant 98-00217, and by the German-Israeli Science Foundation
grant I-629-62.14/1999.


\end{document}